\documentclass[journal=jacsat,manuscript=article]{achemso}

\usepackage{amsmath}
\usepackage{latexsym}
\usepackage{float}
\usepackage{amssymb}
\usepackage{graphicx}
\usepackage{epsfig}
\usepackage{bm}

\newcommand{\x}{{\bm x}}

\newcommand{\be}{\begin{equation}}
\newcommand{\ee}{\end{equation}}
\newcommand{\forget} [1]{}

\author{S. Chibbaro}
\affiliation[Dip. Fisica Universit\'a di Roma Tre Via della vasca navale 84, 00146 Roma, Italy]{Dip. Fisica Universit\'a di Roma tre  Roma, Italy}
\email{sergio.chibbaro@gmail.com}

\author{E. Costa}
\affiliation[D'Appolonia S.p.A. 00142 Rome, Italy]{D'Appolonia S.p.A. 00142 Rome Italy}

\author{D. I. Dimitrov}
\affiliation[Inorganic Chemistry and Physical Chemistry Department University of Food
Technology, Maritza Blvd. 26, 4000 Plovdiv, Bulgaria]{Inorganic Chemistry and Physical Chemistry Department University of Food
Technology Maritza Blvd 26 4000 Plovdiv Bulgaria}

\author{F. Diotallevi}
\affiliation[IAC--CNR, via dei Taurini 19, 00185, Roma, Italy]{IAC--CNR 00185 Roma Italy}

\author{A. Milchev}
\affiliation[Institute of Physical
Chemistry, Bulgarian Academy of Sciences, 1113 Sofia, Bulgaria]{Institute of Physical
Chemistry Bulgarian Academy of Sciences 1113 Sofia Bulgaria}

\author{D. Palmieri}
\affiliation[D'Appolonia S.p.A. 00142 Rome, Italy]{D'Appolonia S.p.A. 00142 Rome Italy}

\author{G. Pontrelli}
\author{S. Succi}
\affiliation[IAC--CNR, va dei Taurini 19, 00185, Roma, Italy]{IAC--CNR 00185 Roma Italy}

\title[Capillary filling with wall corrugations]
{Capillary filling in microchannels with wall corrugations -\\
A comparative study of the Concus-Finn criterion by continuum, kinetic and atomistic approaches}

\begin{document}
\begin{abstract}  We   study  the  impact  of   wall  corrugations  in
microchannels on  the process of  capillary filling by means  of three
broadly   used   methods  -   Computational   Fluid  Dynamics   (CFD),
Lattice-Boltzmann  Equations (LBE)  and Molecular  Dynamics  (MD). The
numerical results of these  approaches are compared and tested against
the Concus-Finn (CF) criterion,  which predicts pinning of the contact
line at  rectangular ridges perpendicular  to flow for  contact angles
$\theta  >  45^\circ$.  While  for  $\theta  =  30^\circ$,  $\theta  =
40^\circ$ (no flow) and $\theta  = 60^\circ$ (flow) all methods are
found to produce  data consistent with the CF  criterion, at $\theta =
50^\circ$ the  numerical experiments  provide different results. Whilst
pinning  of the  liquid  front is  observed  both in  the  LB and  CFD
simulations,  MD simulations  show that  molecular  fluctuations allow
front  propagation even  above  the critical  value  predicted by  the
deterministic CF  criterion, thereby introducing a  sensitivity to the
obstacle height.
\end{abstract}

\section{Introduction}\label{Intro}

In  the recent  years, with  the rapid  technological progress  in the
production  of micro-  and nano-channels,  the understanding  of fluid
flow on the nanoscale\cite{Drake,Meller} has become crucial for modern
nanotechnology (such as the ``lab on a chip'' and related microfluidic
devices) as well as for various applications of porous materials, flow
in biomembranes, etc.  A key problem is the  description of fluid flow
in narrow  channels with wettable walls. Such  channels are ubiquitous
in cells  and living matter  but have been also  successfully produced
from  synthetic materials in  recent years\cite{Alvine}.  Thus, planar
nanochannels, fabricated  by silicon-based technology,  can provide an
attractive   configuration  for   fundamental  studies   like  filling
kinetics\cite{Haneveld},   hydrodynamics  in   confinement,   and  for
molecular separation  processes in biology\cite{Han}.  Indeed, besides
practical applications, microfluidics also raises a challenge to basic
research  since the  continuum description  of fluid  flow  goes under
question whenever discreteness of matter  comes into play, that is, at
length scales comparable to the molecular size.

To gain insight into the transport mechanisms involved in fluid flows,
many researchers  have studied  the  problem using  a  variety of  computer
simulation  methods,  and most  notably  Computational Fluid  Dynamics
(CFD), Lattice-Boltzmann Equations  (LBE), and Molecular Dynamics (MD)
methods. The  classical continuum  theory based on  Navier-Stokes (NS)
equations assumes  that state variables  do not vary appreciably  on a
length scale  comparable to the  molecular free path.  This conjecture
has been challenged by  both experiment\cite{Nagy} and the earliest MD
computer  simulations\cite{Cieplak} which  indicate  the existence  of
significant  density fluctuations of  the liquid  normal to  the solid
wall.      Against       expectations,      however,      some      MD
studies\cite{Bitsanis,Evans} on  Poiseulle flow demonstrated  that the
classical NS  description may be  used for modeling capillary  flow in
channels with  diameter of several molecular sizes  and greater, while
this  has  been challenged  by  another  study\cite{Travis}. To  these
controversial  results  one  should  add  data, produced  by  the  LBE
method\cite{shanchen} which has  gained increased prominence in recent
years due to its efficiency  and proximity to the basic assumptions of
the NS constitutive equations.  Results from the LBE approach indicate
that  there are  several microfluidic  situations, in  which molecular
details, although  non-negligible, can  still be given  a meso-scopic,
rather  than fully  atomistic, representation,  without  affecting the
basic physics  \cite{SBRAG, yeomans1, yeomans2,karlin,dupin,chibbaro,Fabiana}.
On  general grounds,  this can  be expected  to be  the  case whenever
molecular fluctuations  do not play  any major role.  However,  as the
physics of  micro/nanoflows progresses towards  increasingly demanding
standards,  qualitative  expectations  need  to  be  complemented  and
possibly tested against quantitative assessments.

In view of the diversity  of methods and the plethora of controversial
results from the computational  modeling of flow in the sub-micrometer
range,  a  test of  the  adequacy  and  reliability of  the  principal
approaches  is  highly  warranted.  In the  present  investigation  we
perform such  a test by comparing  the results from three  of the most
broadly used methods - CFD, LBE,  and MD - focusing on a generic case,
the  capillary filling  of a  wettable narrow  channel  by spontaneous
imbibition of a simple fluid.

The aim of this comparative study is as follows: (i) we test
the reliability  of the three  simulation methods with respect  to the
capillary filling of a  wettable nanochannel, by comparing the results
against the Lucas - Washburn (LW) theoretical prediction; (ii) we analyse the effect of the presence of an isolated corrugation
on  the flow  dynamics  and  test the  numerical  results against  the
theoretical Concus-Finn criterion;
(iii) the direct comparison between fluctuating (MD) and non-fluctuating (CFD and LB) approaches,
permits to highlight the role of thermal fluctuations on the universality of the CF criterion.

\section{Theoretical  background}  In  a  horizontal  capillary  under
steady state conditions (i.e., in the absence of gravity and transient
inertial effects),  the capillary  imbibition pressure is  balanced by
the viscous drag of the  liquid. Simple analysis of this process leads
to  the  Lucas  -  Washburn equation  (LW)\cite{Lucas,Washburn}  which
relates the distance of penetration  $z(t)$ of the fluid front at time
$t$  to the  capillary  size  $H$, the  viscosity  $\eta$ and  surface
tension $\gamma$ of the liquid,  and the static contact angle $\theta$
between  the liquid  and  the  capillary wall.  If  slip velocity  and
deviations from  Poiseille flow are neglected, in  the late asymptotic
regime the  distance $z(t)$ travelled  by the moving interface  in the
channel (having $z_0$ as initial coordinate) is given by
\begin{equation}\label{LW}   z(t)^2   -   z_0^2   =   \frac{\gamma   H
\cos(\theta)}{\eta} ct,
\end{equation} where  $H$ denotes  the height of  the channel  and the
constant $c=1/3$ for a  slit-like capillary.  Equation \ref{LW} can be
recast    in    dimensionless     form    as    $\tilde{z}=z/H$    and
$\tilde{t}=t/t_{cap}$, where $t_{cap}=H \eta/\gamma$, becoming: \be
\label{LW_dimless}    \tilde{z}(\tilde{t})^2    -   \tilde{z}_0^2    =
\cos(\theta)  c   \tilde{t}.   \ee  It  is  apparent   that  in  these
dimensionless units, time-penetration depends only on the value of the
contact  angle  $\theta$, regardless  of  other  fluid parameters  and
geometrical details. This facilitates the comparison between different
simulation methods.

\begin{figure}[htb]
\begin{center}
\includegraphics[width=6cm,angle=0]{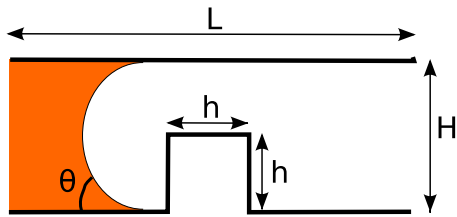}
\caption{\label{fig1} Scheme of the microchannel geometry}
\end{center}
\end{figure}

In this comparative study we explore capillary filling in the presence
of  topological obstacles  (rectangular  ridges) on  the channel  wall
(\ref{fig1}).  Since capillary  filling  is mainly  determined by  the
three-phase  boundary  (``contact  line'')  between liquid,  wall  and
vapor, the  motion of  the contact line  offers a much  more stringent
test of  the various  simulation methods and  offers a  possibility to
assess their shortcomings and advantages. A key problem in this regard
is the pinning of the contact  line, due to a geometric singularity in
the  meniscus stability, like  the presence  of obstacles  whose sides
make an  angle $2\alpha$  with respect to  the wall, broadly  known as
Concus-Finn  condition \cite{Concus,  Concus2, Concus3}.  According to
the  Concus-Finn  criterion,  there  is  no filling  due  to  meniscus
instability,  provided  the   contact  angle  fulfills  the  following
condition:
\begin{equation} \theta > \frac{\pi}{2} - \alpha.
\end{equation} For  a rectangular obstacle, this means  that a contact
line  will be  pinned  at its  trailing  edge, once  $\theta >  45^0$,
regardless of the obstacle  height.  While this condition follows from
the detailed mathematical  analysis\cite{Concus} of the liquid surface
stability, in  fact it goes  back to the thermodynamic  foundations of
wetting, as  established in the early works  of Gibbs\cite{Gibbs}. The
possibility  for capillary  driven  flow is  of  major importance  for
numerous   applications,  e.g.,   modern   fuel  cells\cite{Zengerle},
therefore it  is not  surprising that this  criterion has  been tested
even in space experiments on the Shuttle-Mir complex.

\section{Modelling microfluids: from continuum mechanics to molecular dynamics}\label{model}

Currently, intensive  efforts to get deeper  insight and understanding
of  flow in  microchannels  are  carried out  by  researchers using  a
variety of computer modeling approaches. In principle, continuum fluid
mechanics  provides  the most  economical  description of  microflows.
However, this approach  fails to describe a series  of phenomena, such
as near-wall  slip-flow, in which the continuum  assumption goes under
question due to the onset  of molecular effects, especially near solid
walls. Molecular Dynamics provides a high degree of physical fidelity,
at  the price, however,  of a  very susbtantial  computational burden.
The  lattice kinetic  approach  is  well positioned  to  offer a  good
compromise between the physical  realism of molecular dynamics and the
computational       efficiency       of      continuum       mechanics
\cite{STATPHYS,Gladrow}.   The  lattice  kinetic approach  is  finding
increasing  applications to  microfluidic problems,  as it  permits to
handle fluid-wall  interactions at a  more microscopic level  than the
Navier-Stokes  equations,  while simultaneously  reaching  up to  much
larger            scales            than            Molecular-Dynamics
\cite{SBRAG,Yeom,Harting,Popescu}. However, the LB approach is subject
to a number of limitations, such as the existence of spurious currents
near curved  interfaces, as well  as enhanced evaporation/condensation
effects  due  to the  finite-width  of  the interface  \cite{EPJB09b}.
Ideally,  one would  combine the  three methods  within  a synergistic
multiscale   procedure,   using   each   of   them   wherever/whenever
appropriate, in different regions of the microflow.  Various two-level
(continuum-MD, LB-MD  and continuum-LB), are already  available in the
literature~\cite{Bird},   while,  to  the   best  of   our  knowledge,
three-level  coupling  are just  beginning  to  appear \cite{MU3}.   A
detailed  and comparative assessment  of merits  and downsides  of the
three levels of description, is  therefore of great importance to the
ultimate goal  of developing efficient and  robust multiscale coupling
methods  for  complex microfluidic  flows.   In  this  work, we  shall
present a  comparative study of  capillary filling in the  presence of
wall  corrugations, using all  of the  three methods  mentioned above,
namely a finite-volume CFD software package, a LB model with non-ideal
fluid interactions,  and a  MD computer program.   Since all  of these
methods are by now standard, in  the following we shall provide just a
cursory description,  leaving the details  to the vaste  literature on
the subject.

\subsection{Computational Fluid Dynamics}\label{model_CFD}

Our CFD  approach is based  on the CFD-ACE+ software  package (release
2008), as a multiphysics  commercial tool including geometry modeling,
grid   generation  and   results   visualization  \cite{dapp1}.    The
spontaneous capillary filling in  micro/nano channels is reproduced by
means of the VOF (Volume of  Fluid) scheme, based on the hypotheses of
incompressibility,   no-interpenetration   and  negligible   localized
relative slip of  the two fluids in contact. In  order to describe the
transport of  the volume fraction $\phi$  of one of the  two fluids in
each  cell  ($\phi$ thus  ranging  from  0  to 1),  the  Navier-Stokes
equations for the fluid velocity are augmented with a scalar transport
equation for the volume fraction:
\begin{equation}\label{CFD_eq}  \frac{\partial   \phi}{\partial  t}  +
\vec{\nabla} \cdot (\bm{u} \phi) = 0
\end{equation}  where  $t$ is  time,  $\vec{\nabla}$  is the  standard
spatial gradient operator, and $\bm{u}$  is the velocity vector of the
fluid.  The  composition of the  two fluids in the  mixture determines
for each computational cell  the averaged value of physical properties
such  as  density  and  viscosity.  Any  volume-specific  quantity  is
evaluated in accordance with the following expression:
\begin{equation} \bar \omega=\phi \omega_l + (1-\phi) \omega_g
\end{equation}  where $\bar \omega$  is the  volume-averaged quantity,
$\omega_l$ (resp.   $\omega_g$) is the  value of the property  for one
liquid  (resp.   gas).   However,  being $\phi$  the  averaged  volume
fraction  of fluid  in  each  cell, the  definition  of the  interface
between  the  two fluids  in  that  cell is  not  unique  and must  be
dynamically  reconstructed throughout  the  simulation.  In  CFD-ACE+,
this  is accomplished through  an upwind  scheme with  PLIC (Piecewise
Linear Interface Construction  method \cite{dapp2,dapp3}), taking into
account  surface   tension  to  determine   accurately  the  interface
curvature.   As to  boundary  conditions, a  zero  static pressure  is
imposed at  both inlet  (liquid only) and  outlet (gas only).   At the
initial time  both fluids  are at  rest, with a  short portion  of the
channel at the inlet  filled with liquid.  This specific configuration
allows   the   liquid-vapour  interface   to   assume  the   curvature
corresponding to the hydrophilic  partial wetting condition imposed at
walls  through  the  contact   angle  $\theta$,  thus  generating  the
capillary pressure  that drives the fluid motion.   The CFD simulation
set-up consists of a $2D$  straight channel with a height $H=800\; $nm
and overall length $L=80\; \mu$m.  At the bottom wall, a squared post,
of side and height $h=400\; $nm, is located.  The computational domain
has  been  discretized  with  a squared  structured  non-uniform  grid
consisting of $185000$ cells and $190000$ nodes.

\subsection{Lattice      Boltzmann      method     for      multiphase
flows}\label{geom_LBE}

\begin{figure}[htb]
\includegraphics[width=3cm,angle=0]{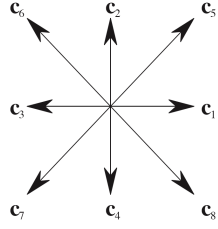}
\caption{\label{d2q9} The two-dimensional, nine-speed LB scheme.}
\end{figure}

The Lattice  Boltzmann (LB) method is  based on a minimal  form of the
Boltzmann kinetic  equation describing the time  evolution of discrete
particle   distribution   function   $f_i(\bm{x},t)$,   denoting   the
probability of  finding a particle  at lattice site $\bm{x}$  and time
$t$ moving  along the lattice  vector $\bm{c}_i$ (  \ref{d2q9}), where
the index $i$ labels the discrete directions of motion.

In  mathematical  terms  \cite{Gladrow},  the  LB  equation  reads  as
follows: \cite{SS_92} \be
\label{eq:LB}         f_{i}(\bm{x}+\bm{c}_{i}\Delta         t,t+\Delta
t)-f_i(\bm{x},t)=-\frac{\Delta                           t}{\tau}\left(
f_{i}(\bm{x},t)-f_{i}^{(eq)}(\rho,\rho{\bm u}) \right) + F_i \; \Delta
t  \ee where  $\rho$ is  the  fluid density,  ${\bm u}$  is the  fluid
velocity and $i=0-8$ labels the the nine-speed, two-dimensional $2DQ9$
model  \cite{Gladrow}. The second  term on  the right  hand side  is a
model collision operator $\Omega_i$, describing the relaxation towards
a local equilibrium on a  time scale $\tau$.  Finally, $F_i$ describes
the  effect  of  external/internal  forces on  the  discrete  particle
distribution.  The  macroscopic density  and momentum are  obtained as
weighted averages of discrete distributions: \cite{Gladrow}
\begin{equation}  \rho(\x,t)=\sum_{i}  f_{i}(\x,t);  \qquad \rho  {\bm
u}(\x,t)=\sum_{i}{\bm c}_{i}f_{i}(\x,t).\end{equation} For the case of
a two-phase fluid, the interparticle force reads as follows: \be
\label{forcing} \bm{F}=-{\cal  G} c^{2}_{s}\sum_i w_i  \psi(\x,t) \psi
(\x+{\bm c}_i\Delta t,t) \bm{c}_i \quad i=0-8.  \ee

Here  $\psi(\x,t)=\psi(\rho(\x,t))=(1-\exp(-\rho(\bm  x,t)))$  is  the
pseudo-potential  functional, describing the  fluid-fluid interactions
triggered by inhomogeneities of the density profile, and ${\cal G}$ is
the  strength   of  the  coupling   (see  \cite{shanchen,prenoi2}  for
details).  Finally,  $F_i=w_i \;{\bm F} \cdot  {\bm c}_i/c_s^2$, where
$w_i=0,1/9,1/36$   are  the   standard  weights   of   the  nine-speed
two-dimensional lattice 2DQ9 \cite{Gladrow} and $c_s=1/\sqrt 3$ is the
sound speed in lattice  units.  Besides introducing a non-ideal excess
pressure $p^{\star}=\frac{1}{2} {\cal G} c_s^2 \psi^2$, the model also
provides a  surface tension $\gamma \sim {\cal  G} c_s^4 \frac{(\delta
\psi)^2}{\delta_w}$,  where   $\delta  \psi$   is  the  drop   of  the
pseudo-potential across the interface  of width $\delta_w$.  By tuning
the  value of the  pseudo-potential at  the wall  $\psi(\rho_w)$, this
approach allows  to define a  static contact angle, spanning  the full
range of values $\theta \in [0^o:180^o]$ \cite{Kang}.

As  to  the boundary  conditions,  the  standard  bounce-back rule  is
imposed: any  flux of particles  that hits a boundary  simply reverses
its  velocity  so  that  the  average  velocity  at  the  boundary  is
automatically zero.  This rule can  be shown to yield no-slip boundary
conditions  up  to   second  order  in  the  Knudsen   number  in  the
hydrodynamic limit of single phase flows.

\begin{figure}[htb]
\begin{center}
\includegraphics[width=8cm,angle=0]{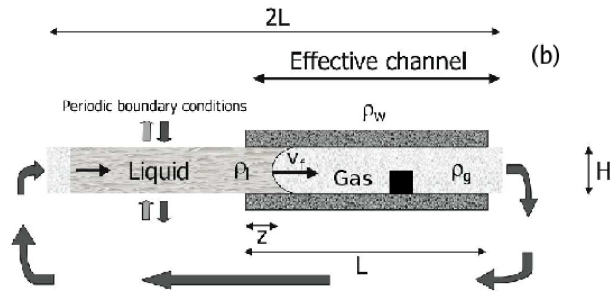}
\caption{\label{fig_fab} Geometrical set-up of the LB simulations.}
\end{center}
\end{figure}

In  this work  we consider  a $2D$  channel composed  of  two parallel
plates separated by  a distance $H= 40 \Delta$,  where $\Delta$ is the
space discretization unit.  This two-dimensional geometry, with length
$2  L=  3000   \Delta$,  is  divided  in  two   regions,  as  shown  in
\ref{fig_fab}.   The left part  has top  and bottom  periodic boundary
conditions, so  as to support  a perfectly flat  gas-liquid interface,
mimicking  a  ``infinite reservoir''.   The  actual capillary  channel
resides on the right half, of length $L$.  The top and bottom boundary
conditions  are those  of a  solid wall,  with a  given  contact angle
$\theta$.  Periodic  boundary conditions are also imposed  at the west
and east  sides. The squared  obstacle, of side $h=H/2=20  \Delta$, is
placed on the lower wall.

\subsection{Molecular Dynamics}
 
During the last few decades Molecular Dynamics has proved to be one of
the most  broadly used  simulation techniques, capable  of reproducing
many details  of macroscopic fluid  dynamics. It has  been extensively
used to study  dynamic wetting problems and to shed  more light on the
molecular     processes     close      to     the     contact     line
region~\cite{Robins,Gentner}.   An  appealing  feature  of  MD  (e.g.,
compared with Monte Carlo or molecular statics) is that it follows the
actual dynamical evolution of the system.

We perform MD simulation of a simple generic model on a coarse-grained
level. In our MD simulation,  see Fig. 4, the fluid particles interact
with each other through a Lennard-Jones (LJ) potential,
\begin{equation} U_{LJ}(r)=4\epsilon[(\sigma/r)^{12} -(\sigma/r)^{6}]
\end{equation} where  $\epsilon=1.4$ and $\sigma=1.0$.   The capillary
walls are  represented by particles forming a  triangular lattice with
spacing $1.0$ in units of  the liquid atom diameter $\sigma$. The wall
atoms may  fluctuate around their equilibrium positions,  subject to a
finitely extensible non-linear elastic (FENE) potential
\begin{equation} U_{FENE}(r) = -15 \epsilon_w R_0^2\ln(1-r^2/R_0^2)
\end{equation}  with $R_0  = 1.5$  and $\epsilon_w  = 1.0  k_BT$ where
$k_B$ denotes the Boltzmann constant and $T$ is the temperature of the
system; $r$ is the distance between the particle and the virtual point
which represents  its equilibrium position in the  wall structure. The
FENE-potential acts like an  elastic string between the wall-particles
and their equilibrium positions in the lattice and keeps the structure
of the wall densely-packed  hexagonal. In addition, the wall particles
interact with each other via a LJ potential with $\epsilon_{ww} = 1.0$
and $\sigma_{ww}  = 0.8$.  This  choice of interactions  guarantees no
penetration  of liquid  through the  wall, and  at the  same  time the
mobility of the wall particles corresponds to the system temperature.

Molecules are advanced in  time via the velocity-Verlet algorithm with
integration time step  $\delta t = 0.01 t_0$  where $t_0= (\sigma^2m /
48 \epsilon_{LJ})^{1/2}  = 1/\sqrt{48}$ is the basic  time-unit and we
have  taken particle  mass  $m=1$ and  $k_BT=1$.   The temperature  is
maintained by  a Dissipative-Particle-Dynamics (DPD)  thermostat, with
friction parameter $\xi  = 0.5$, thermostat cutoff $r_c  = 2.5 \sigma$
and   step-function-like  weight  functions   \cite{HK,Allen}.   Fluid
properties and flow boundary conditions arise then as a consequence of
the collision  dynamics and the  local friction controlled by  the DPD
thermostat. An advantage of the DPD method in comparison with other MD
schemes  is  that  the  local  momentum  is  conserved,  so  that  the
hydrodynamic  behavior of  the  liquid at  large  scales is  correctly
reproduced.

It is worth emphasizing that  all contact angles $\theta$, used in the
simulations, have been determined by measuring the mean static contact
angle $\theta$ of a sessile meniscus in a slit, formed by two parallel
atomistic   walls  for   a   few  given   liquid  -wall   interactions
$\epsilon$. Specific  contact angles are then  chosen by interpolation
between  the relevant  values  of $\epsilon$.   Therefore, the  actual
value  of the  contact angle  when the  fluid front  is located  at an
obstacle edge, for instance, may incidentally deviate from the nominal
value of  $\theta$ . A  recent MD study~\cite{Dimitrov_PRL}  has shown
that  the  LW-law  (Eq.(1)),   holds  almost  quantitatively  down  to
nanoscale tube diameters. In this study we use an obstacle shaped as a
rectangular ridge  which runs perpendicular  to the flow  direction in
the  slit and  has the  same atomic  composition as  that of  the slit
walls. The  height and width  of the ridge  are chosen as  $10 \sigma$
 so that the obstacle height takes half of the slit thickness.

\section{Numerical              results}\label{results} 
We  have  performed simulations  of
spontaneous capillary  filling in presence of a  squared obstacle, for
each  of the three  computational methods  described above.   The test
case  is the  same: the  filling of  a channel  of height  $H$  with a
squared post of side $h=H/2$.  In LB and MD, the post is placed at the
central position on the bottom wall(\ref{fig1}), at a distance $l=25H$
and $l=5H$  from the inlet, respectively, due  to computational costs.
Since the LW  equation does not depend on the  channel length, this is
not a  limiting factor for the  present study (see  below).  Since the
inertial  transient  with the  CFD  method  is  much longer  than  the
estimated inertial time, as shown also in \ref{CFD_snap}, in this case
the channel length has been taken $l \approx 100H$, so as to guarantee
that the front meets the post when the asymptotic regime is attained.
 
We  have performed  a series  of four  simulations (for  each method),
varying the contact angle  $\theta$ from $30^\circ$ to $64^\circ$. The
specific values of the  physical parameters for each simulation method
are reported in Table 1.

\begin{table}
\begin{center}
\begin{tabular}{|l|l|l|l|} \hline \em Parameter  & \em CFD value & \em
LB value & \em MD value\\\hline

Channel  height $H$  (m) &  $8 \cdot  10^{-7}$ &  $8 \cdot  10^{-7}$ &
$8\cdot 10^{-7}$\\

Water density $\rho_l$ (kg/m$^3$)& $10^3$ &$10^3$ &$10^3$ \\

Water kinematic  viscosity $\nu$ (m$^2$/s)  & $10^{-6}$ &  $10^{-6}$ &
$10^{-6}$\\

Water  dynamic viscosity  $\eta$  (kg/(m s))&  $10^{-3}$  & $10^{-3}$  &
$10^{-3}$ \\

Liquid/gas surface  tension $\gamma$ (N/m)& $7.2  \cdot 10^{-2}$ &$9.6
\cdot 10^{-2}$ & $3.4 \cdot 10^{-4}$\\

Vapour  density  $\rho_g$ (kg/m$^3)$&  $1.167$  &  $28.12$ &$10.6$  \\
\hline

\end{tabular}
\caption{Physical  quantities for  the three  different  cases.  Units
have been chosen as follows: LB) $\Delta x=H/NY$, $\Delta t=\nu_{LB}\;
\Delta  x^2 /\nu$, $\Delta  m= \rho_l\;  \Delta x^3  /\rho_{LB}$); MD)
$\sigma = H/20$, $\tau=  \nu_{MD} \; \sigma^2/\nu$, $\delta m=\rho_l\;
\sigma^3  /\rho_{MD}$. The  different values  of the  gas  density and
surface tension reflect the computational constraints of the different
methods.}
\end{center}
\end{table}

Some comments are in order.  By construction, the CFD approach is able
to reproduce exactly all the  physical properties of the flow.  On the
other hand, both LB and MD show some discrepancies in the value of the
surface tension and vapour density.  However, these discrepancies have
been  found to  be  irrelevant  to the  purpose  of investigating  the
macroscopic              features             of             capillary
imbibition~\cite{Dimitrov_PRL,EPJB09}

\begin{figure}[htb]
\includegraphics[scale=0.6]{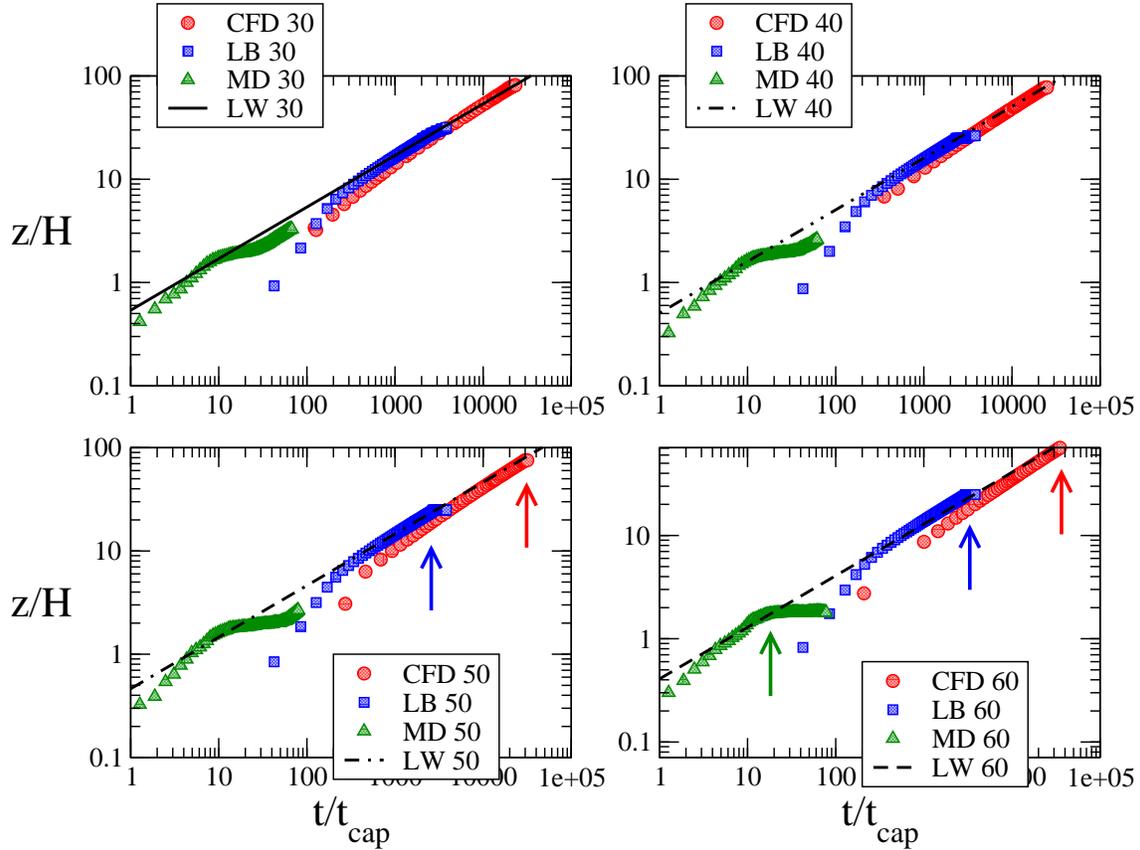}
\caption{
\label{all}  Log-Log  plot   of  the  dimensionless  front  coordinate
$z(t)/H$ vs  the dimensionless time $t/t_{cap}$. Here  $z$ denotes the
centerline position  ($y=H/2$) of the front, $y$  being the cross-flow
coordinate.  All simulations show superposition with the LW-prediction
before reaching  the obstacle, at  all inspected contact  angles.  The
arrows  indicate the  pinning points  for the  case of  $50^\circ$ and
$60^\circ$,  not  visible on  the  scale  of  this figure.   See  also
\ref{CFD_snap}, \ref{LBE_snap} and \ref{MD_snap} for a close-up of the
dynamics around the obstacle region only. }
\end{figure}

As  is  well  known, depending  on  the  value  of the  contact  angle
$\theta$, two scenarios  may appear according to the  Gibbs, or Concus
Finn criterion  \cite{Concus, Concus2, Concus3}: a)  for small contact
angles  the front  is able  to  climb the  obstacle, walk  on it,  and
eventually pass it;  b) for large contact angles  the front climbs the
obstacle, but pins  at its back edge, thus  stopping the fluid motion.
We wish to emphasize that the Concus-Finn, or Gibbs, criterion is
based  upon thermodynamic  arguments  and, consequently,  it has  been
rigourously proven only at a macroscopic level.

\subsection{Recovering the Lucas-Washburn regime}

As is  well known, the  Lucas-Washburn asymptotic regime sets  in once
inertial effects die out and  steady propagation results from the sole
balance between capillary drive and dissipation.  During the transient
regime, the  dynamic contact angle,  which depends on the  front speed
itself,  changes   in  time  until   its  static  value   is  attained
\cite{Martic,Quere}.

In  \ref{all}, the time  evolution of  the centerline  ($y=H/2)$ front
position  is reported as  a function  of time,  and compared  with the
dimensionless  LW-law Eq.   (\ref{LW_dimless}).  As  one can  see, all
three methods  exhibit satisfactory agreement with  the theoretical LW
prediction, although on a different range of time-scales.  This is due
to the  different values  of the parameters  used in each  method, and
particularly,   to   the   fact    that   the   MD   capillary   speed
$V_{cap}=\gamma/\eta$ is $20 \div 30)$  times higher than in the other
methods.  Indeed, it  should be noted that the  LW asymptotic solution
sets in  after a  typical transient  time of the  order of  $\tau \sim
H^2/12  \nu_l$,  or  in  units  of capillary  times,  $\tau/t_{cap}  =
\frac{\rho \gamma  H}{12 \mu^2}$.  Based on  the data in  Table I, one
can  readily check  that $\tau/t_{cap}  \sim 1$  for LB  and  CFD, and
$\tau/t_{cap}  \sim 10^{-2}$  for  MD.  It  is  therefore possible  to
appreciate  the anomalous transient  in the  CFD case,  see \ref{all},
which is of the order $\tau/t_{cap} \sim 100$.  However, since CFD and
LB simulations  last over $10^3$  capillary times, and  MD simulations
last  about $10$ capillary  times, before  reaching the  obstacle, the
superposition with the LW regime is free from transient phenomena.
 
\subsection{Front morphology while crossing the obstacle}

\begin{figure}[htb]
\includegraphics[scale=0.4]{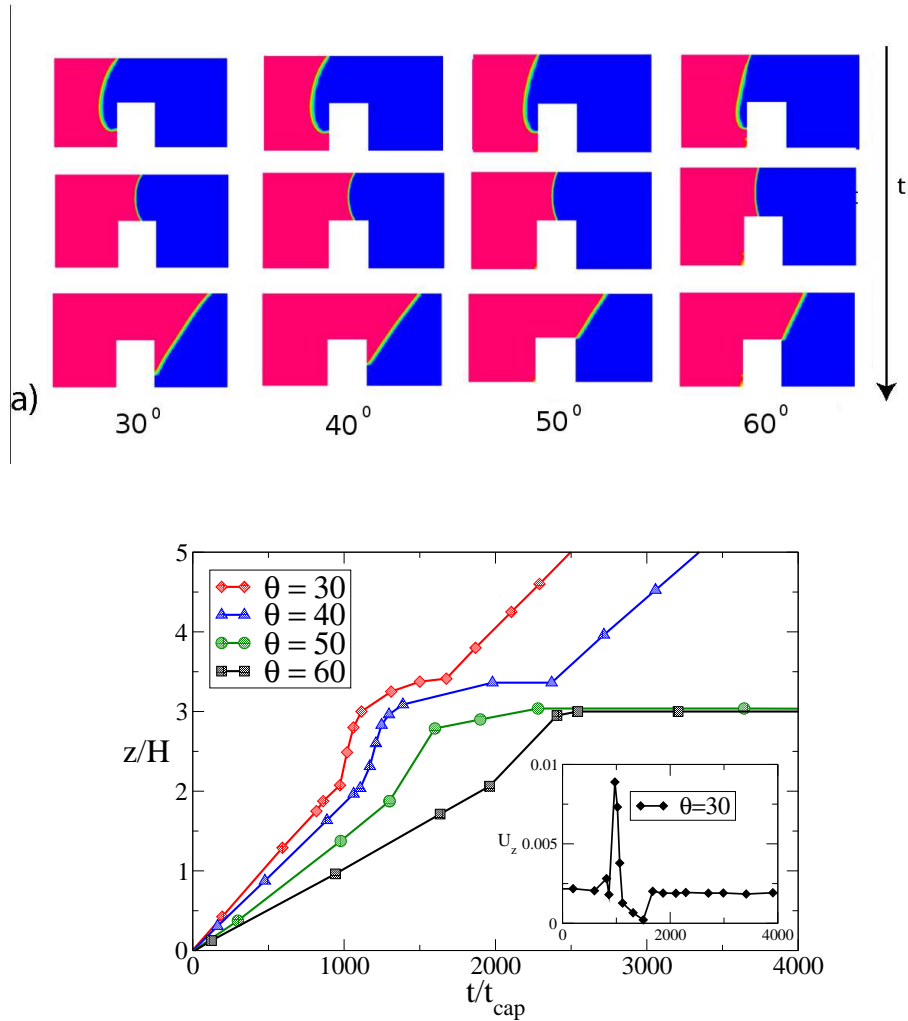}       \vspace{1cm}
\hspace{1cm}
\includegraphics[scale=0.35]{CFD_data_new.eps}
\caption{\label{CFD_snap}  a)  Snapshots  of  the  front  dynamics  at
different stages  of a post  surmounting CFD simulation.   The various
columns  from left  to right  correspond to  contact angles  $\theta =
30^{\circ},\; 40^{\circ},\;50^{\circ},\;60^{\circ}$,  as indicated. b)
Time evolution  of the interface midpoint while  crossing the obstacle
in  CFD simulations.   The inset  shows the  instantaneous propagation
speed of the front for the case $\theta=30^{\circ}$.  }
\end{figure}

\begin{figure}[htb]
\includegraphics[scale=0.5]{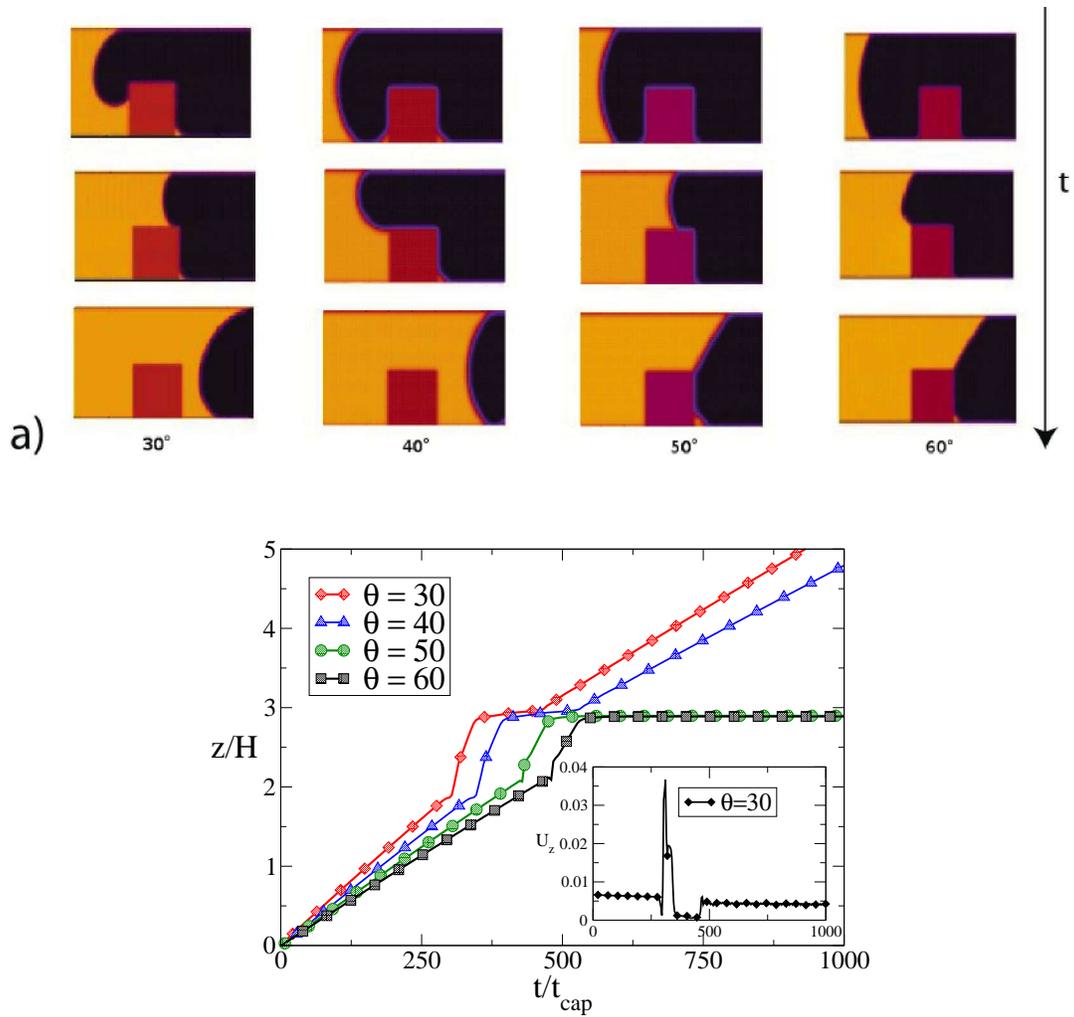}\\
\vspace{1truecm}
\includegraphics[scale=0.35]{LB_data_new.eps}
\caption{\label{LBE_snap}  a)  Snapshots  of  the  front  dynamics  at
different stages  of a  post surmounting LB  simulation. From  left to
right,   the  columns   correspond   to  contact   angles  $\theta   =
30^{\circ},\;    40^{\circ},\;50^{\circ},\;60^{\circ}$.     b)    Time
evolution of the interface midpoint  while crossing the obstacle in LB
simulations.  The  inset shows the instantaneous  propagation speed of
the front for the case $\theta=30^{\circ}$.}
\end{figure}

\begin{figure}[htb]
\includegraphics[scale=0.8]{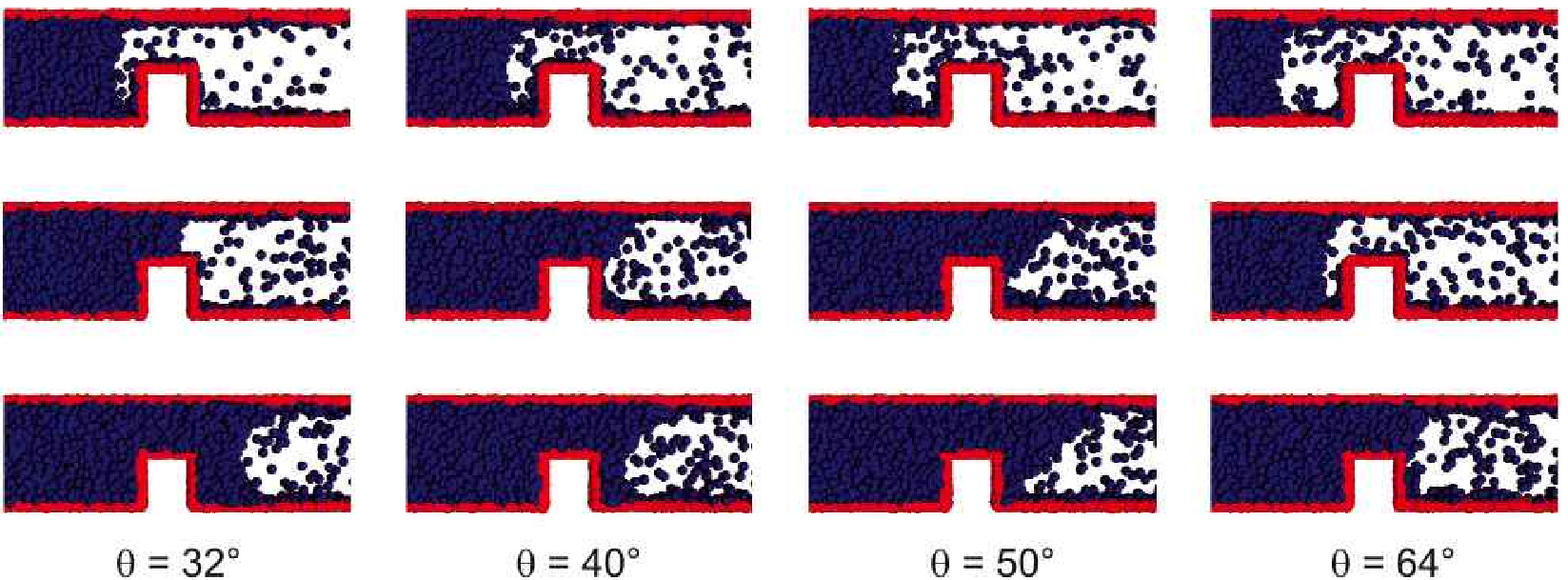}\vspace{1cm}
\\
\includegraphics[scale=0.3]{MD_data_new.eps}
\caption{\label{MD_snap} a) Variation of the number of fluid particles
$N$  in the  capillary  with time  $t$  for contact  angles $\theta  =
32^{\circ},\; 40^{\circ},\;50^{\circ},\;64^{\circ}$. Symbols represent
MD results,  whereas lines denote the predicted  behavior according to
the  LW-equation.   The   vertical  grey-shaded  column  indicates  an
extension  in the  time  axis.   b) Time  evolution  of the  interface,
computed through the fluid volume measurement,
  while crossing the  obstacle in  MD simulations.   The inset
shows the instantaneous  propagation speed of the front  for the cases
$\theta=32^{\circ}$.  }
\end{figure}

In \ref{CFD_snap}, \ref{LBE_snap}  and \ref{MD_snap}, snapshots of the
fluid front  during the surmounting  of the obstacle are  reported for
the three  simulation methods  (the figures report  density contours).
As  one  can see,  the  front  dynamics  and morphology  are  strongly
affected by the presence of  the obstacle: the liquid impinging on the
obstacle must  adjust to a  $90^{\circ}$ degrees discontinuity  of the
contact angle,  which is clearly causing a  significant deformation of
the front, before the static  value is recovered again on the flat-top
surface of the  obstacle.  These changes of shape  are well visible in
figures \ref{CFD_snap}  and \ref{LBE_snap}, for both CFD  and LB.  The
case of  MD is less  clear-cut, due to  the absence of  a well-defined
interface and  to molecular fluctuations.  Indeed,  although the fluid
meniscus is  clearly visible, its  surface appears rough  and strongly
fluctuating in  time. From \ref{MD_snap}  one can see that  some atoms
evaporate  from the  liquid  and overcome  the  slit.  Moreover,  long
before the fluid meniscus  has passed the obstacle, vapor condensation
and partial filling of the wedge, formed by the rear wall of the ridge
and the slit  wall, take place (see also  \ref{MD_snap}a). Since fluid
imbibition in  a capillary  is accompanied by  the faster motion  of a
precursor film far ahead of the meniscus \cite{Bonn,Kav_03,sergio}, it
is also conceivable  that the wedge is filled  by this atomically thin
precursor. Since  the meniscus  position at time  $t$ is  difficult to
locate precisely,  we rather  measure the volume  of the fluid  in the
capillary  which is  readily obtained  by  the total  number of  fluid
atoms, residing  at time $t$ in  the slit during  the filling process.
For an  incompressible fluid, this volume is  directly proportional to
the distance  $z(t)$, travelled  by the fluid  front at time  $t$. The
curve giving $z(t)$ is shown  in ~\ref{MD_snap}b for several values of
the contact  angle, $30^{\circ} \le \theta \le  64^{\circ}$. The shape
changes also entail  a substantial change of the  front speed, as well
documented  in \ref{CFD_snap}b,  \ref{LBE_snap}b and  \ref{MD_snap}b ,
from which the front speed can be  read off simply as the slope of the
curves  reporting the  front position  as a  function of  time.  These
figures  show evidence  of  a significant  acceleration  of the  front
(centerline  position $y=H/2$) in  the climbing  stage, followed  by a
deceleration  in the  stage where  the front  is approaching  the rear
corner, where the chance/risk of pinning is highest.  Here the fate of
the  front becomes  critical.  According  to the  CF analysis,  if the
contact angle  is below $45^{\circ}$,  there is enough drive  from the
upper  wall  to  pull  the  front  away from  the  rear  edge  of  the
obstacle. Otherwise, the front stops moving.

The CFD and  LB simulations confirm this picture,  showing evidence of
pinning only  for the two angles above  $45^{\circ}$.  More precisely,
they  show a  significant front  acceleration in  the  climbing stage,
followed by  a coasting  period, once the  rear edge is  reached.  For
$\theta=30^{\circ}$  and   $\theta=40^{\circ}$,  the  coasting  period
exhibits a finite lifetime, after  which the front regains its motion.
For $\theta=50^{\circ}$  and $\theta=60^{\circ}$, the  coasting period
does not seem to come to  an end (within the simulation time), and the
front is pinned.   After the obstacle, the (unpinned)  front is slowed
down, as  one can  appreciate from the  slightly reduced slope  of the
front  dynamics,  see   \ref{CFD_snap}b  (only  for  $\theta=40$)  and
\ref{LBE_snap}b.

The MD simulations, on the other hand, tell a different story, in that
even at $\theta = 50^{\circ}$,  the front proves capable of overcoming
the obstacle,  although with a drastically reduced  speed.  Note that,
while both CFD and LB show  evidence of a strong front acceleration as
they approach the obstacle,  MD simply shows a monotonic deceleration.
This  is due  to the  fact that  while in  CFD and  LB  simulations we
measure the  distance travelled by the interface  midpoint, as already
pointed  out,  MD  measures the  total  volume  of  the fluid  in  the
capillary.

Given  the  qualitatively  different  outcome  of  CFD(LB)  versus  MD
simulations at  $\theta > 45^{\circ}$, we next  discuss the dependence
of the filling dynamics on the different contact angles, case by case.
 
\begin{itemize}
\item{Contact angles $30^{\circ}$ and $40^{\circ}$} \label{3040}\\
  As already  mentioned, in  these cases, all  three methods  give the
same qualitative outcome: in proximity of the ridge, the front deforms
in response  to the geometrical discontinuity, climbs  up the obstacle
and walks over its top. Once  the rear-edge is reached, the bottom end
of  the  front  pins at  the  corner,  and  keeps  moving on  the  top
wall. Then,  according to  the CF criterion,  the front  overcomes the
obstacle and completely fills  up the entire channel.  Manifestly, the
standard  $z$  versus  $t$   relationship,  described  by  the  LW-law
(\ref{LW})  is strongly  violated  in the  vicinity  of the  obstacle.
After  overcoming the  obstacle,  the usual  LW  regime is  recovered,
although after  a transitional  period of time,  which depends  on the
wettability  $\theta$~\cite{chibbaro}, and  with  a reduced  velocity.
In particular, in all cases, the front is slowed down right after
passing  the   obstacle,  see  \ref{CFD_snap}b,   \ref{LBE_snap}b  and
\ref{MD_snap}b.  However, at  $\theta=30$, the asymptotic behaviour is
quite  different  for  the  three  methods: (i)  In  CFD  simulations,
\ref{CFD_snap}b, after a short  transient time, the velocity basically
regains the initial value before passing the obstacle. (ii) LB shows a
similar behaviour, with only a  slight reduction of the front velocity
after passing the obstacle. (iii)  in MD, however, the front undergoes
a substantial velocity decrease.

This might  be due to the details  of the fluid-solid interaction
during the obstacle  surmounting (in which the contact  angle is bound
to   undergo  drastic   changes),  as   well  as   to   the  different
time-scales. A detailed analysis of this effect shall be deferred to a
future study. 

\item{Contact  angle $50^{\circ}$}  \label{50} \\In  this  case, while
results from CFD and LB  confirm the CF criterion, MD simulations show
a  different  behaviour  for  the  front  motion:  interestingly,  the
Concus-Finn (Gibbs) criterion for contact  line pinning at the edge of
the ridge is  found to break down. Indeed,  for $\theta = 50^{\circ}$,
the fluid  front overcomes the  obstacle in manifest violation  of the
Concus-Finn criterion.  In order to vividly explain this feature,
we show in \ref{FigMD4050} the snapshots and in video 1 and  2 the movie of the front dynamics, respectively
for the  cases $\theta =  40^{\circ}, ~50^{\circ}$.  These observations
suggest  that,  on  the   nanoscale,  the  overcoming  of  topological
obstacles  is  strongly   affected  by  interface  fluctuations,  thus
undermining    the   deterministic    nature    of   the    imbibition
process.  Indeed,  both the  CFD  and LB  method  work  in absence  of
fluctuations,  and   this  would   explain  the  difference   with  MD
simulations.  The problem  of  contact line  pinning during  capillary
imbibition acquires  thus a stochastic character and  is most probably
governed by  the size of the  obstacles around the  CF critical point.
In fact, depending on the  height of the ridge obstacle, a coalescence
of the  pinned meniscus with the  molecules ahead of  the obstacle, in
the vicinity of the edge, may occur at later times.

\item{Contact angle  $60^{\circ}$ }\label{60} \\Again, in  this case all
the three methods give the  same result: the front deforms, climbs the
obstacle, walks on  its top, but pins at the  back edge and definitely
stops moving.

\begin{figure}[htb]
\includegraphics[scale=1.2]{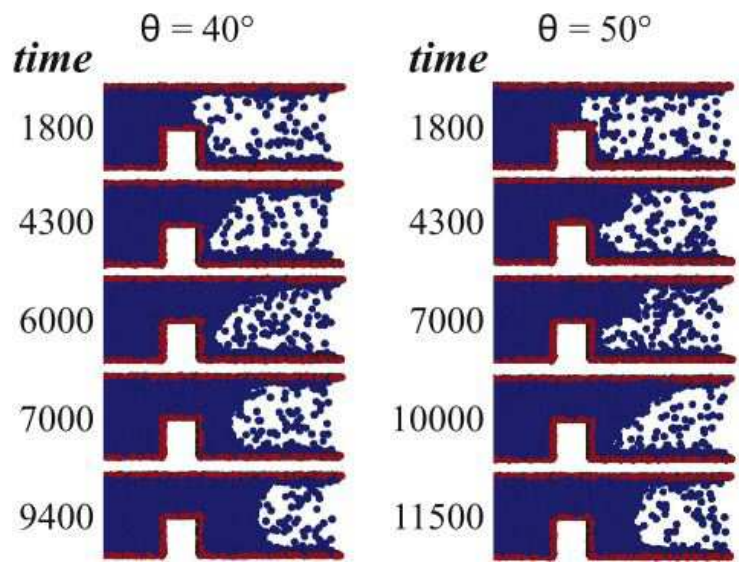}
\caption{\label{FigMD4050}   Snapshots from MD simulations with $\theta=40^{\circ}$ (on the left) and 
for $\theta=50^{\circ}$ (on the right). It is clearly seen that for both values of the contact angle the front overcomes the obstacle, although at different times. This shows that the Concus-Finn criterion is violated in the case with $\theta=50^{\circ}$.}
\end{figure}  

\subsection{Further discussion}

In order to  gain a better understanding of  the previous results, and
particularly  of  the  violation   of  the  CF  criterion  for  mildly
super-critical  angles in  MD simulations,  additional runs  have been
performed.    More  precisely,   we   have  run   MD  simulations   at
$\theta=50^{\circ}$ with a taller obstacle, $h=15 \sigma$, in order to
inspect whether the front is still capable of surmounting the obstacle
in violation of the CF criterion  (in this case, the total slit height
was correspondingly increased, so as to leave the same clearance above
the obstacle as in the  previous simulations).  Moreover, we have also
run  additional CFD  and  MD simulations  at $\theta=50^{\circ}$  with
shorter obstacles,  $h=H/4$, in order to inspect  whether even CFD(LB)
would show violations  of the CF criterion upon  reducing the obstacle
size.   The main  outcome is  as  follows: MD  simulations with  $h=15
\sigma$ {\it do} show front pinning, indicating that violations of the
CF  criterion disappear once  the obstacle  is made  sufficiently tall
(see Fig.  \ref{FigMDnew}).  This corroborates our previous conjecture
of a (Arrehnius-like?) dependence of  the CF criterion on the obstacle
height, in the presence of  molecular fluctuations.  At the same time,
CFD  and  LB  simulations  with $h=H/4$  at  $\theta=50^{\circ}$  keep
showing evidence  of pinning, thereby  lending further support  to the
idea that  the CF criterion  remains insensitive to obstacle  size, so
long as molecular fluctuations can be neglected.

\begin{figure}[htb]
\includegraphics[scale=1.0]{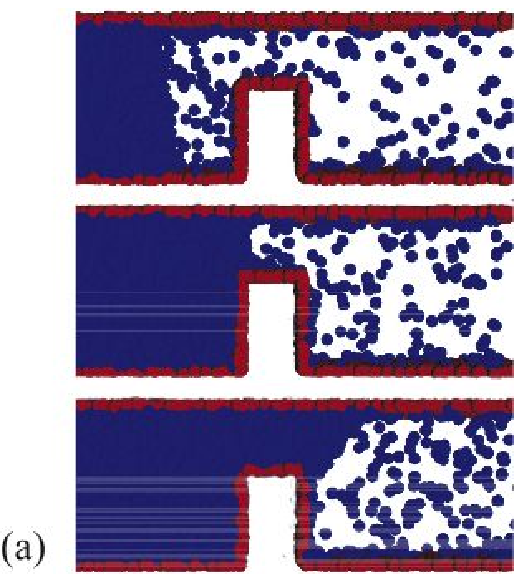}
\includegraphics[scale=1.]{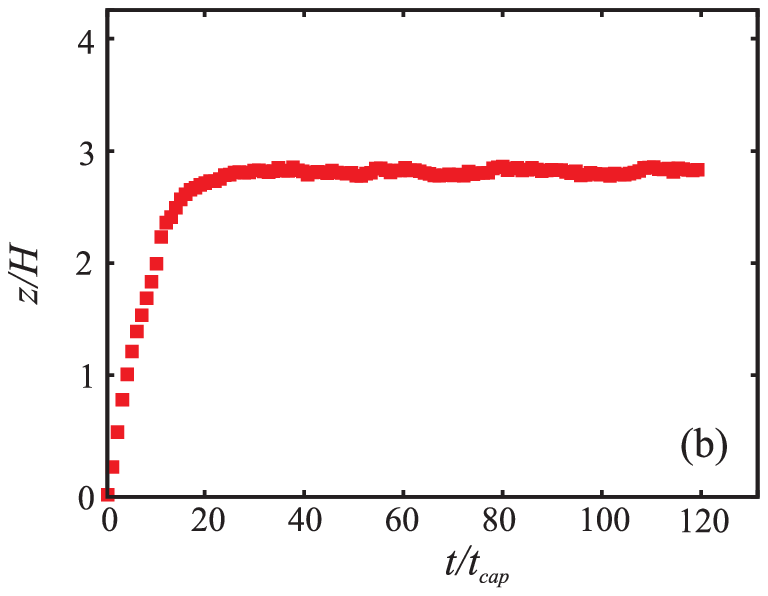}
\caption{\label{FigMDnew}   Snapshots   (a)   and   centerline   front
coordinate (b), from MD simulations with $\theta=50^{\circ}$ and $h=15
\sigma$.  Unlike  the same case with  $h=10 \sigma$, the  front is now
pinned in accordance with the CF criterion.  }
\end{figure}  

Before concluding, a few words of  caution should be spent on the fact
that,  being diffuse-interface methods,  both CFD  and LB  propagate a
finite-width interface.   As a result,  one may wonder whether  and to
what extent  finite-width effects could interfere with  the physics of
post surmounting.  Indeed,  as shown in previous studies  by these and
other   authors   (\cite{EPJB09,yeomans1}),   finite   interface   width
eventually leads to mild deviations from the LW law.  However, they do
not  affect  the qualitative  outcome  of  a pinning/no-pinning  test,
unless the  interface width  becomes comparable to  the characteristic
obstacle width.  Both CFD and LB simulations are visibly far from this
critical limit,  which is  why we are  confident that  the qualitative
conclusions  of the  present work  are not  significantly  affected by
finite-width   effects.  
One may wonder whether such deviations might be paralleled to the effect of molecular
fluctuations. To this regard, it is worth underlining that such a 
parallel has to be taken very cautiously, since molecular fluctuations stem from 
the microscopic physics of the problem, while finite-width effects are due to a lack
of numerical resolution. Sometimes a mapping between the two can be established, but this
can by no means be taken as a general rule.
Finally,   another  potential   source   of
discrepancy between CFD(LB) and MD  is the fact that the latter allows
slip motion  while CFD and LB (in  this set up) do  not.  Although any
solid statement  on the effect  of slip flow  on the dynamics  of post
surmounting must  necessarily be  deferred to a  detailed quantitative
analysis,  we  observe that  due  to  the weakness  of  inertial
effects, hydrodynamic boundary conditions have little or no effect, on
the dynamics/energetics of the obstacle surmounting.

This is confirmed  by the fact that slip  flow effects should manifest
through visible  deviations from  the LW regime,  of which we  have no
evidence, at least in the  parameter regime investigated in this work.
This  situation  can  drastically  change  in  the  presence  of
superhydrophobic  effects, although  we  shall not  be concerned  with
these problems in the present work.

\end{itemize}
\section{Conclusions}\label{summary} Summarizing,  we have studied the
effect  of geometrical obstacles  in microchannels  on the  process of
capillary  filling, by means  of three  distinct simulation  methods -
Computational Fluid Dynamics  (CFD), Lattice-Boltzmann Equations (LBE)
and Molecular Dynamics (MD). The numerical results of these approaches
have been compared and  tested against the Concus-Finn (CF) criterion,
which  predicts pinning  of  the contact  line  at rectangular  ridges
perpendicular to  flow for contact  angles $\theta >  45^\circ$. While
for  $\theta =  30^\circ$ (flow),  $\theta =  40^\circ$ and  $\theta =
60^\circ$ (no flow)  all methods are found to  produce data consistent
with  the   CF  criterion,  at  $\theta  =   50^\circ$  the  numerical
experiments provide  different outcomes.  While pinning  of the liquid
front is observed  in both LB and CFD  simulations, the MD simulations
show that the  moving meniscus overcomes the obstacle  and the filling
goes  on, for a sufficiently small obstacle.  This result  indicates that  the  macroscopic picture
underlying  the CF  criterion and  hydrodynamic approach  needs  to be
amended near  the critical  angle.  Furthermore, while  in CFD  and LB
simulations the  front re-emerges from the obstace  surmounting with a
nearly  unchanged  velocity,  in  the  MD  case  the  post-surmounting
velocity appears considerably reduced.   These results suggest that,
away from the critical value $\theta=45^o$, the issue of front-pinning
in a corrugated  channel can be quantitatively described  by a kinetic
Boltzmann approach or by the macroscopic CFD method.

 While the CFD software used in this work is well-suited to handle
complex  geometries, it  also  shows some  physical and  computational
limitations, namely anomalous-long transients  and the need of a large
computational grid  to assure the  required accuracy, which  entails a
correspondingly  long  computational  time,  much  closer  to  the  MD
requirements than  to the LB ones.   In the vicinity  of the critical
angle,  the motion  of  the  front exhibits  a  strong sensitivity  to
molecular  fluctuations  which cannot  be  accounted  for by  standard
(non-fluctuating)  LB  methods,   let  alone  continuum  methods.   In
particular, the MD simulations  show that molecular fluctuations allow
front propagation  slightly above the critical value  predicted by the
deterministic CF  criterion, thereby introducing a  sensitivity to the
obstacle height  (the CF criterion  is restored for  sufficiently tall
obstacles).   On the basis of the present results, it would be indeed of interest to explore whether fluctuating
hydrodynamic methods, either in the form of stochastic hydrodynamics or fluctuating LB, would 
prove capable of reproducing the results of MD simulations~\cite{Landman}.
 Whether  the   probability  of   "tunnelling"   to  the
deterministically    forbidden   region    $\theta>45^o$    shows   an
Arrehnius-like  dependence  on  the  obstacle  height,  stands  as  an
interesting topic for future research.

\begin{acknowledgement} The authors thank the support from the project
"INFLUS", NMP-031980 of the VI-th FW programme of the EC.
\end{acknowledgement}


\end{document}